\documentclass[twocolumn]{aastex631}

\usepackage{xspace}
\usepackage{booktabs} 
\usepackage{multirow} 

\newcommand\Teff{$T_{\rm{eff}}$}
\newcommand\logg{$\log g$}
\newcommand\logKzz{$\log K_{zz}$}
\newcommand\lmix{$l_{\rm{mix}}$}
\newcommand\um{$\mu$m\xspace}

\begin{document}

\title{Diversity of Cold Worlds: Predicted Near- to Mid-infrared Spectral Signatures of a Cold Brown Dwarf with Potential Auroral Heating}

\author[0000-0002-2011-4924]{Genaro Su\'arez}
\affiliation{Department of Astrophysics, American Museum of Natural History, Central Park West at 79th Street, NY 10024, USA}
\correspondingauthor{Genaro Su\'arez}
\email{gsuarez@amnh.org}

\author[0000-0001-6251-0573]{Jacqueline K. Faherty}
\affiliation{Department of Astrophysics, American Museum of Natural History, Central Park West at 79th Street, NY 10024, USA}

\author[0000-0003-4600-5627]{Ben Burningham}
\affiliation{Department of Physics, Astronomy and Mathematics, University of Hertfordshire, Hatfield, UK}

\author[0000-0002-4404-0456]{Caroline V. Morley}
\affiliation{University of Texas at Austin, 2515 Speedway, Austin, TX 78712}

\author[0000-0003-0489-1528]{Johanna M. Vos}
\affiliation{School of Physics, Trinity College Dublin, The University of Dublin, Dublin, Ireland}

\author[0000-0002-9420-4455]{Brianna Lacy}
\altaffiliation{51 Pegasi b Fellow}
\affil{Department of Astronomy, University of California Santa Cruz}
\affil{Bay Area Environmental Research Institute + NASA Ames Research Center}

\author[0000-0003-4225-6314]{Melanie J. Rowland}
\affil{University of Texas at Austin, Department of Astronomy, 2515 Speedway C1400, Austin, TX 78712, USA}

\author[0000-0002-6294-5937]{Adam C. Schneider}
\affil{United States Naval Observatory, Flagstaff Station, 10391 West Naval Observatory Road, Flagstaff, AZ 86005, USA}

\author[0000-0003-0548-0093]{Sherelyn Alejandro Merchan}
\affiliation{Department of Astrophysics, American Museum of Natural History, Central Park West at 79th Street, NY 10024, USA}
\affiliation{Department of Physics, Graduate Center, City University of New York, 365 5th Ave., New York, NY 10016, USA}

\author[0000-0001-8170-7072]{Daniella C. Bardalez Gagliuffi}
\affiliation{Department of Physics \& Astronomy, Amherst College, Amherst, MA, USA}

\author[0000-0003-2235-761X]{Thomas P. Bickle}
\affil {School of Physical Sciences, The Open University, Milton Keynes, MK7 6AA, UK}

\author[0000-0003-4636-6676]{Eileen C. Gonzales}
\affiliation{Department of Physics \& Astronomy, San Francisco State University, 1600 Holloway Ave., San Francisco, CA 94132, USA}

\author[0000-0003-2102-3159]{Rocio Kiman}
\affil{Department of Astronomy, California Institute of Technology, Pasadena, CA 91125, USA}

\author[0000-0003-4083-9962]{Austin Rothermich}
\affiliation{Department of Astrophysics, American Museum of Natural History, Central Park West at 79th Street, NY 10024, USA}
\affiliation{Department of Physics, Graduate Center, City University of New York, 365 5th Ave., New York, NY 10016, USA}
\affiliation{Department of Physics and Astronomy, Hunter College, City University of New York, 695 Park Avenue, New York, NY, 10065, USA}
\affiliation{Backyard Worlds: Planet 9}

\author[0000-0001-8818-1544]{Niall Whiteford}
\affiliation{Department of Astrophysics, American Museum of Natural History, Central Park West at 79th Street, NY 10024, USA}




\begin{abstract}

Recent JWST/NIRSpec observations have revealed strong methane emission at 3.326~\um in the $\approx$482~K brown dwarf CWISEP~J193518.59$-$154620.3 (W1935). Atmospheric modeling suggests the presence of a $\approx$300~K thermal inversion in its upper atmosphere, potentially driven by auroral activity. We present an extension of the retrieved spectra of W1935 with and without inversion spanning 1--20~\um, to identify thermal inversion-sensitive spectral features and explore the origin of the object's peculiar characteristics. Our analysis indicates that atmospheric heating contributes approximately 15\% to the bolometric luminosity. The model with inversion predicts an additional similar-strength methane emission feature at 7.7~\um and tentative ammonia emission features in the mid-infrared. Wavelengths beyond $\sim$2~\um are significantly influenced by the inversion, except for the 4.1--5.0~\um CO$_2$ and CO features that originate from atmospheric layers deeper than the region where the inversion occurs. W1935 appears as an outlier in Spitzer/IRAC mid-infrared color-magnitude diagrams (CMDs) based on the $m_{\rm Ch1}-m_{\rm Ch2}$ (IRAC 3.6~\um $-$ 4.5~\um) color, but exhibits average behavior in all other combinations  that trace clear sequences. This anomaly is likely due to the Ch2 filter probing vertical mixing-sensitive CO$_2$ and CO features that do not correlate with temperature or spectral type. We find that the thermal inversion tends to produce bluer $m_{\rm Ch1}-m_{\rm Ch2}$ colors, so the overluminous and/or redder position of W1935 in diagrams involving this color cannot be explained by the thermal inversion. This analysis provides insights into the intriguing dispersion of cold brown dwarfs in mid-infrared CMDs and sheds light on their spectral diversity.
\end{abstract}



\section{Introduction}
Auroral activity has been detected on Earth and all four gas giants in the Solar System \citep[e.g.,][]{Broadfoot_etal1979,Broadfoot_etal1981,Broadfoot_etal1989,Bhardwaj_Gladstone2000}. 
This phenomenon arises from interactions between electrically charged particles and the neutral upper atmosphere as the particles travel along magnetic field lines. 
Auroral activity has also been observed on Venus and Mars \citep{Phillips_etal1986,Bertaux_etal2005}, which do not possess internal magnetic fields, highlighting that auroral phenomena can be driven by alternative mechanisms. 
Aurorae, therefore, appear to be common in our Solar System. While most of these auroral emissions have been detected at radio and ultraviolet wavelengths \citep[e.g.,][]{Lamy2020}, 
near-infrared detections have also been reported using data from the Juno mission \citep[e.g.;][]{Mura_etal2017} and ground-based telescopes (e.g. Gemini North; \citealt{Kedziora-Chudczer_etal2017} and Keck; \citealt{Thomas_etal2023}).

Brown dwarfs also exhibit auroral activity, detected primarily through radio observations \citep[e.g.,][]{Kao_etal2016,Kao_etal2018,Guirado_etal2025}. 
Efforts to identify corresponding infrared features from the ground have found null detections \citep{Gibbs_Fitzgerald2022}. 
Recently, \citet{Faherty_etal2024} used the James Webb Space Telescope (JWST) to report an infrared feature tentatively attributed to being driven by auroral activity in a brown dwarf without irradiation from a host star. 
They detected a methane emission feature at 3.326~$\mu$m in JWST NIRSpec $R\sim2700$ spectra of CWISEP J193518.59–154620.3 \citep[W1935;][]{Marocco_etal2019}, a brown dwarf with an effective temperature of $\approx$482~K \citep{Faherty_etal2024} and a spectral type of $\ge$Y1 \citep{Meisner_etal2020}. 
To model this feature, the authors performed a retrieval analysis and found that a thermal inversion of approximately 300~K in the upper atmosphere, centered at 1--10~mbar, is required to reproduce the observed emission (see Figure 3 in \citealt{Faherty_etal2024}). 

In addition to producing distinct emission features, auroral activity can influence the continuum from X-ray to infrared wavelengths \citep{Hallinan_etal2015}, potentially driving photometric variability. This is particularly compelling as a likely mechanism contributing to the variability observed in the coldest brown dwarfs \citep[e.g.,][]{Cushing_etal2016,Esplin_etal2016}, where silicate clouds in their photospheres (commonly invoked for warmer brown dwarfs e.g.; \citealt{Radigan_etal2014}) cannot explain the observed variability.  
In contrast to the warmer L and early-T dwarfs, where variability is typically attributed to inhomogeneous silicate clouds \citep[e.g.,][]{Burgasser_etal2002,Radigan2014}, dust clouds in late-T and Y dwarfs are expected to have settled below the visible atmosphere, making silicate cloud-driven variability less likely. However, lighter condensates composed of salts and sulfides, for example, are expected to form in the atmospheres of these cold objects \citep{Morley_etal2012}, which could also induce flux variations. 
Additionally, auroral phenomena are also linked to rotation, with rapid rotation playing a key role in enhancing radio auroral emissions \citep{Kao_etal2018}. 
Therefore, understanding the effects of auroral processes in substellar atmospheres is essential for interpreting observations of cold brown dwarfs and gas giant planets and for identifying the dominant mechanism driving their variability.

In this paper, we aim to identify thermal inversion-sensitive spectral features in near- to mid-infrared atmospheric models and evaluate their role in shaping the spectra of cold brown dwarfs and gas giant planets. 
We describe the observations and models in Section~\ref{sec:data}. In Section~\ref{sec:results}, we identify and discuss the primary spectral changes induced by heating in the upper atmosphere, examine how they influence mid-infrared color–magnitude diagrams, and evaluate the performance of self-consistent atmospheric models to reproduce the data. We summarize our findings and conclude in Section~\ref{sec:conclusions}.

\section{Data}
\label{sec:data}
The observed and modeled near- to mid-infrared spectrophotometry for W1935 described in this section are shown in Figure~\ref{fig:SED}.

\subsection{Observations}\label{sec:observations}
The only existing spectroscopic observations for W1935 were presented in \citet{Faherty_etal2024}. The data were acquired with JWST NIRSpec G395H as part of Cycle 1 GO program 2124 (PI. Faherty). 
The final G395H spectrum has a wavelength coverage from 2.86 to 5.13~$\mu$m and a resolving power of $R\sim2700$ with a median signal-to-noise ratio (S/N) of 16 in the entire range or 30 for detector NRS2 ($>3.79~\mu$m). 
This program also obtained MIRI F1000W, F1280W, and F1800W broadband photometry centered at 9.9, 12.8, and 17.9~$\mu$m, respectively.

Additional photometry for W1935 includes Gemini North Near-InfraRed Imager (NIRI) $J_{\rm{MKO}}$ magnitude (23.93$\pm$0.33~mag) \citep{Leggett_etal2021}, Spitzer IRAC Ch1 (18.51$\pm$0.03~mag) and Ch2 (15.53$\pm$0.02~mag) magnitudes \citep{Meisner_etal2020}, and WISE W1 (18.754$\pm$0.289~mag) and W2 (15.792$\pm$0.061~mag) magnitudes \citep{Marocco_etal2021}.

A reference object observed as part of the same JWST GO program 2124 and also presented in \citet{Faherty_etal2024} is WISE J222055.31$-$362817.4 (W2220 for short). 
This object has very similar physical parameters (temperature, luminosity, radius, and mass) to those of W1935 \citep{Faherty_etal2024}, making it an ideal point of comparison. However, they lie at different places in the $M_{\rm Ch2}$ vs. $m_{\rm Ch1}-m_{\rm Ch2}$ diagram. While W1935 looks overluminous in IRAC Ch2 for its $m_{\rm Ch1}-m_{\rm Ch2}$ color, W2220 has an average IRAC Ch2 magnitude (Section~\ref{sec:CMDs}). 
Both W2220 and W1935 NIRSpec G395H spectra are shown in Figure~\ref{fig:SED_with_W2220}. 
They exhibit strong CH$_4$, but W2220 does not display the 3.326~\um methane emission feature. They also show similar CO, but slightly different CO$_2$ features.
We use here the NIRSpec G395H spectrum and the MIRI photometry of W2220 as a comparison for W1935. 

\subsection{Models}
The W1935 NIRSpec spectrum was modeled in \citet{Faherty_etal2024} using the \textit{Brewster} retrieval framework \citep{Burningham_etal2017,Burningham_etal2021}. 
The best fit retrieved spectrum includes a thermal inversion of $\approx$300~K in the upper atmosphere ($\lesssim$0.3~bar) required to reproduce the strong 3.326~$\mu$m methane emission feature observed in the data. 
The authors speculate that a plausible explanation for this inversion is heating by auroral activity. 
We note, however, that while the retrieval analysis is grounded in local thermodynamic equilibrium (LTE), auroral processes are non-LTE phenomena.  
A comparison retrieval where a thermal inversion was not allowed was also presented in \citet{Faherty_etal2024}. 
We extended both of those retrieved models (with and without a thermal inversion) into the near- and mid-infrared (1--20~\um) to investigate their spectral properties and assess the impact of the inversion.
We note that \citet{Faherty_etal2024} presented only the models within the G395H coverage (2.86--5.13~\um). 
Furthermore, no spectroscopic observations beyond 5.13~$\mu$m have been obtained for W1935. 

To provide context for the W1935 predictions, we also consider the best retrieved spectrum of the W2220 G395H spectrum obtained by \citet{Faherty_etal2024} (Figure~\ref{fig:SED_with_W2220}). Unlike the best retrieved spectrum for W1935, the best model for W2220 does not need a thermal inversion in the atmosphere to successfully reproduce its G395H spectrum (see Extended Data Figure 3 and Extended Data Table 4 in \citealt{Faherty_etal2024}).

\section{Results}
\label{sec:results}

\begin{figure*}
	\centering
	\includegraphics[width=0.9\linewidth]{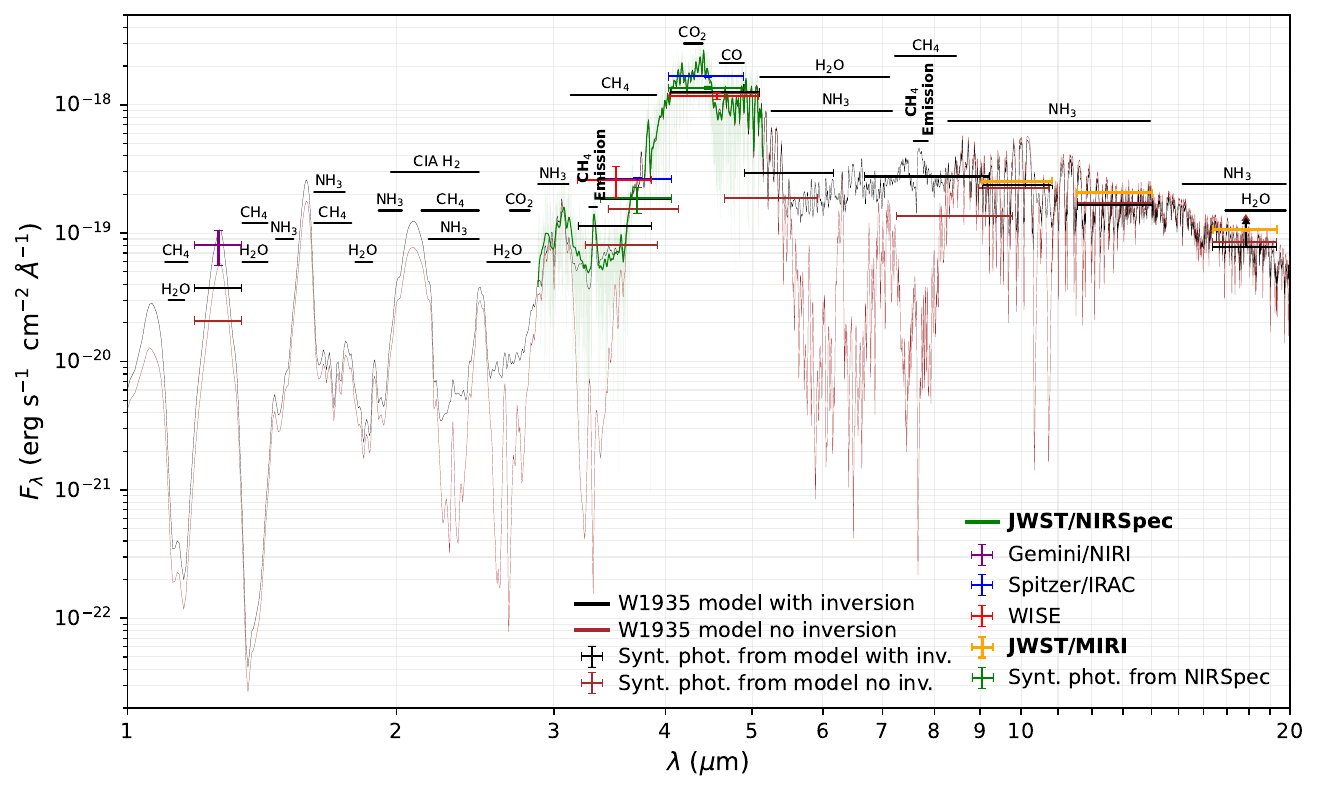}
	\caption{Observed and predicted SEDs of W1935. The observations include the NIRSpec G395H spectrum (original $R\sim2700$ resolution in light green and $R\sim250$ convolved spectrum in dark green) and NIRI, IRAC, WISE, and JWST/MIRI magnitudes (Section~\ref{sec:observations}), as indicated in the bottom-right legend. 
    The best retrieved spectra in \citet{Faherty_etal2024} using either a thermal inversion (black line) or no thermal inversion (brown line) in the upper atmosphere are shown. 
    Synthetic fluxes from both retrieved spectra for the filters with observations, along with IRAC Ch3 (5.6~\um) and Ch4 (7.6~\um) are plotted using the same color as their corresponding spectra. Additional synthetic IRAC Ch1 and Ch2 fluxes from the NIRSpec spectrum are shown in green. The main spectral features are indicated (see Figure~\ref{fig:SED_CF}), including the principal methane emission regions.}
	\label{fig:SED}
\end{figure*}

\begin{figure*}
	\centering
	\includegraphics[width=0.9\linewidth]{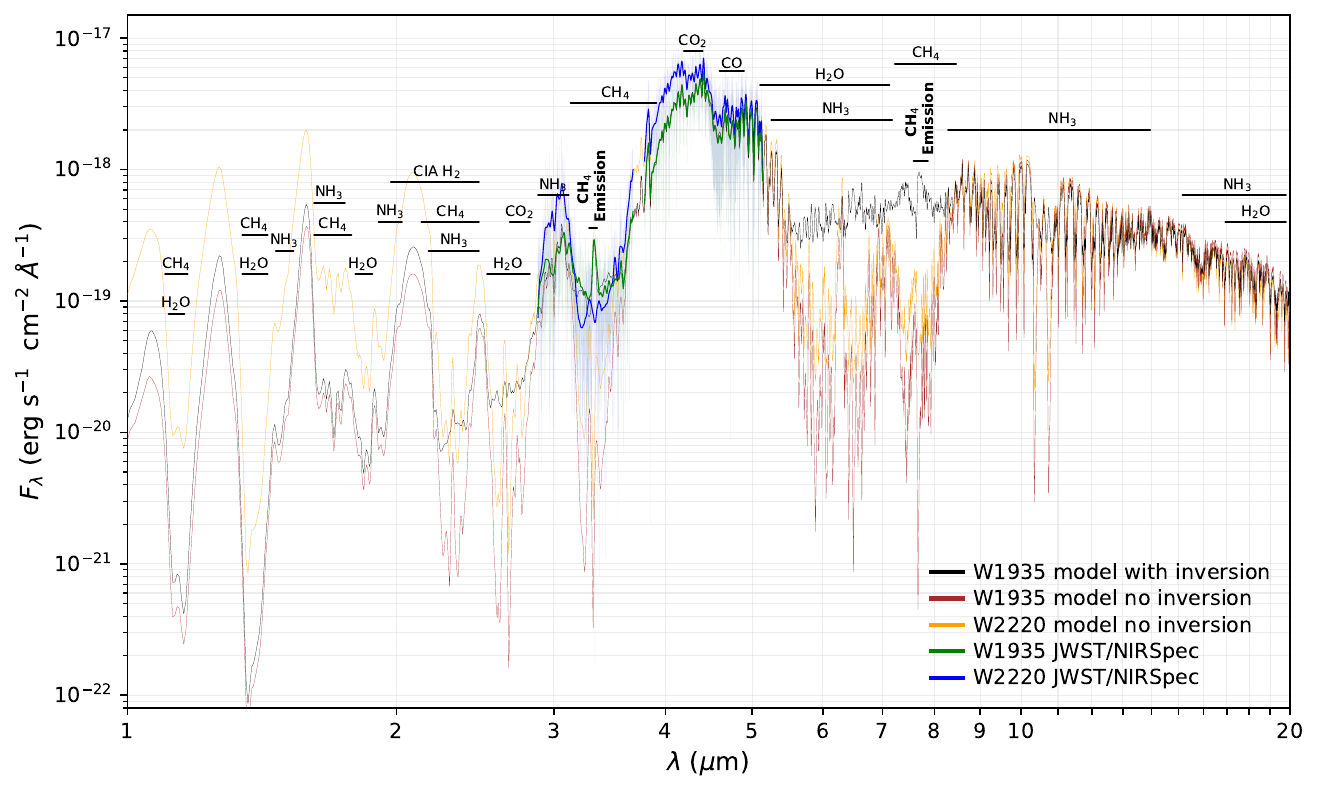}
	\caption{Retrieved spectra for W1935 (with and without thermal inversion in black and brown, respectively) and W2220 (orange curve; no temperature inversion). The NIRSpec G395H spectra for W1935 and W2220 are shown in green and blue, respectively, with a light color for the original resolution ($R\sim2700$) spectra and a darker color for convolved ($R\sim100$) spectra. To facilitate comparison between the two objects, we plotted absolute fluxes, accounting for their respective distances (14.43$\pm$0.79~pc and 10.47$\pm$0.23~pc for W1935 and W2220, respectively; \citealt{Kirkpatrick_etal2021}).}
	\label{fig:SED_with_W2220}
\end{figure*}

\begin{figure*}
	\centering
	\includegraphics[width=1.0\linewidth]{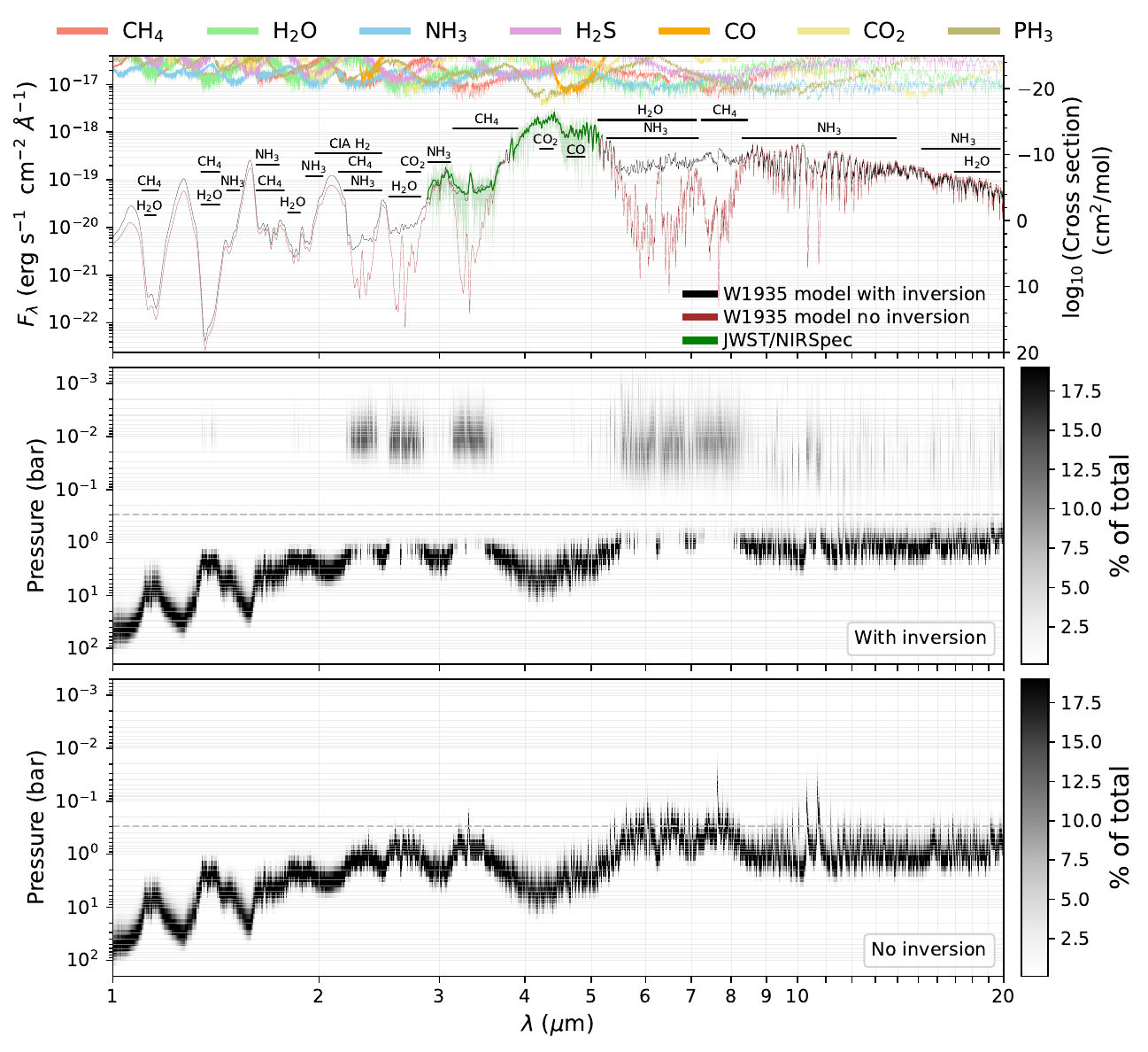}
	\caption{\textbf{Top panel:} 
    SED for W1935 in Figure~\ref{fig:SED} along with the absorption cross sections for key molecular species at a representative temperature of 475~K and a pressure of 1~bar. \textbf{Middle panel:} Contribution function for the retrieved spectrum with thermal inversion. The horizontal dashed line indicates the approximate maximum pressure at which the thermal inversion occurs. \textbf{Bottom panel:} Contribution function for the retrieved spectrum without thermal inversion. The horizontal dashed line indicates is the same as in the middle panel.}
	\label{fig:SED_CF}
\end{figure*}

\begin{figure*}
	\centering	\includegraphics[width=0.7\linewidth]{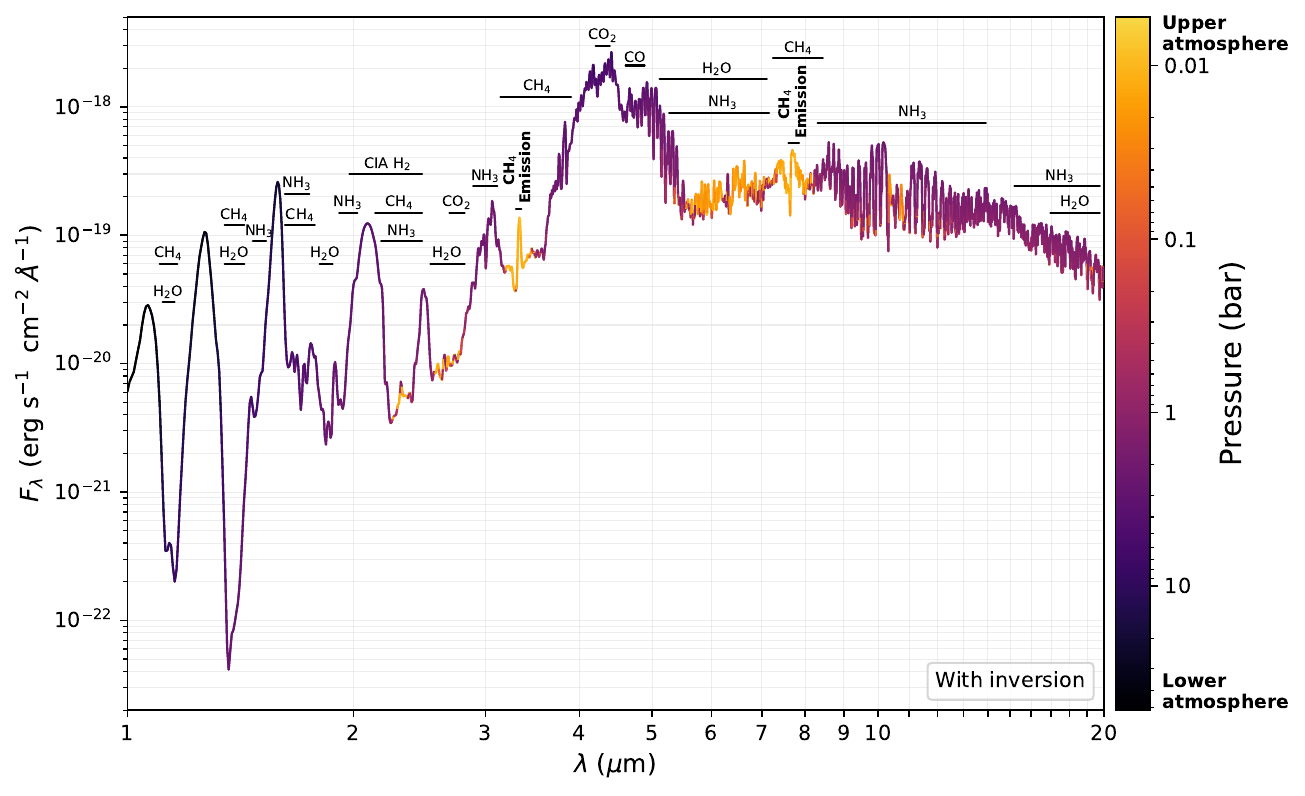}	\includegraphics[width=0.7\linewidth]{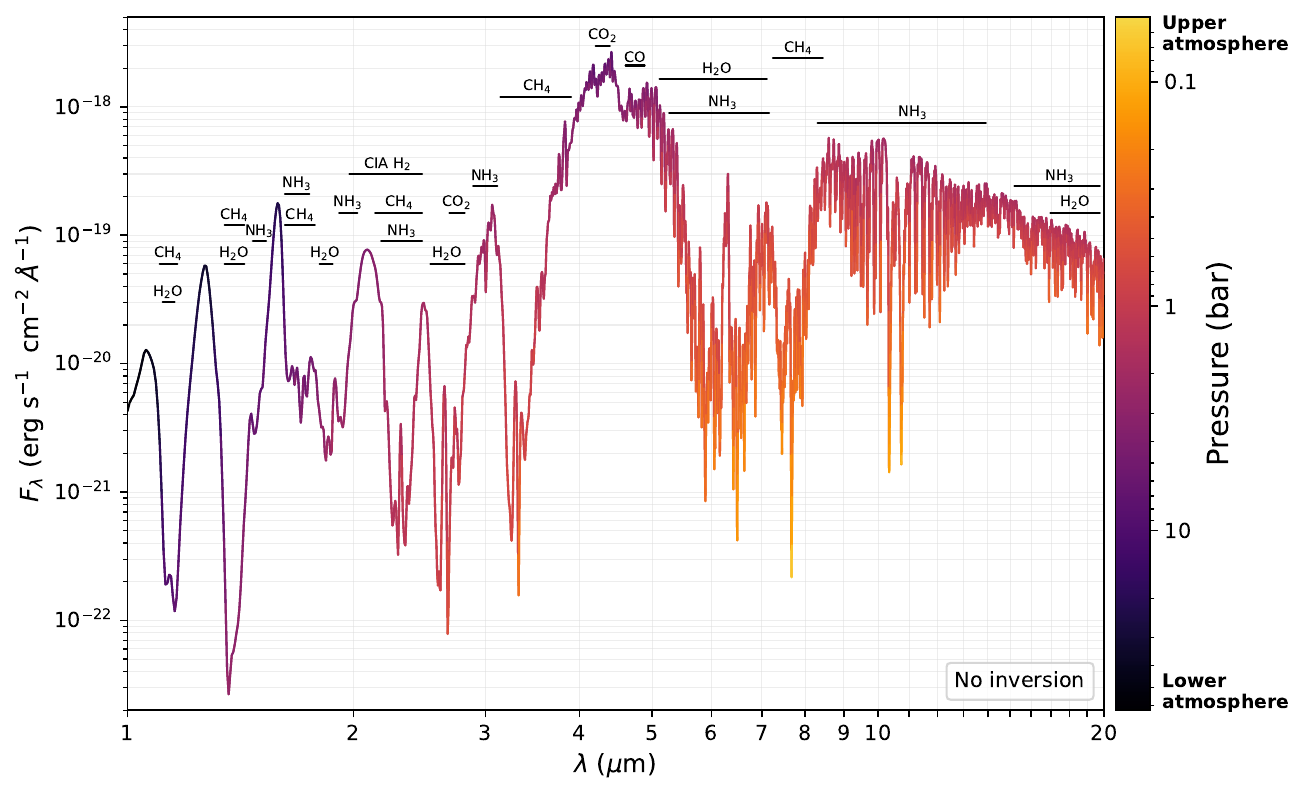}
	\caption{Best retrieved spectra for W1935 with (top panel) and without (bottom panel) thermal inversion. The color axis in each panel shows the corresponding contribution function, which indicates the pressure contributing the most to the fluxes emerging at different wavelengths. 
    The maximum pressure probed by both retrieved spectra is similar ($\approx$60~bar). However, due to the thermal inversion, the minimum pressure probed by the retrieved spectrum with inversion ($\approx$0.002~bar) is much lower that the one from the model without inversion ($\approx$0.1~bar, with a few wavelength points reaching as low as 0.03~bar). The W1935 2.9--5.1~\um G395H spectrum probed pressures $\approx$0.002--10~bar. The main spectral features are indicated, including the principal methane emission regions.}
	\label{fig:SED_aux}
\end{figure*}

Figure~\ref{fig:SED} shows the observed and modeled spectral energy distributions (SEDs) of W1935 
(Section~\ref{sec:data}) from 1~\um up to 20~$\micron$.
We derived synthetic photometry for W1935 from both the best retrieved spectra with and without inversion for the filters for which we have imaging magnitudes (Section~\ref{sec:observations}) using the \texttt{SEDA} tool \citep{Suarez_etal2021a}. 
In addition to these filters, we also considered IRAC Ch3 (5.6~\um) and Ch4 (7.6~\um) because they cover an interesting region of the SED discussed in Section~\ref{sec:predicted_SED}.
Additional synthetic IRAC Ch1 and Ch2 fluxes were computed using the W1935 NIRSpec spectrum. 
We filled in a 3.69--3.79~$\micron$ gap in the NIRSpec spectrum---produced by the fixed slit mode with both NRS1 and NRS2 NIRSpec detectors--- by linearly interpolating the median fluxes in a $\sim0.01~\mu$m window (20 data points) right before and after the gap.
We did not obtain synthetic WISE W1 and W2 photometry from the NIRSpec spectrum because a fraction of the filter transmission curves is not covered by the spectrum ($\approx$4\% for W1 and $\approx$7\% for W2).

\subsection{Luminosity Change}
\label{sec:Lbol_change}
We estimated the bolometric luminosity for W1935 and W2220 by integrating the modeled SEDs over the full 1--20~\um range (Figure~\ref{fig:SED_with_W2220}). We obtained $\log (L_{\rm bol}/L_\odot)$ values of $-6.54^{+0.03}_{-0.02}$ for W1935 and $-6.49^{+0.01}_{-0.01}$ for W2220 from the best models for each object.
These values are slightly fainter than the estimates in \citet{Faherty_etal2024} ($-6.3\pm0.1$ and $-6.4\pm0.1$ for W1935 and W2220, respectively), but consistent within $2.5\sigma$ for W1935 and $1\sigma$ for W2220. 
We note that these differences in luminosity could be attributed to the different approaches used.
While we integrated the 1--20~\um modeled SEDs, \citet{Faherty_etal2024} integrated the full observed SEDs assembled by combining all available data for these objects, which are based on photometric points in the mid-infrared.  
Both sets of $\log (L_{\rm bol})$ estimates indicate that W2220 is $\approx$15--20\% fainter than W1935.

To assess the impact of atmospheric heating on the bolometric luminosity of W1935, we also calculated the luminosity using the retrieved spectrum without inversion. 
This yielded a luminosity of $\log (L_{\rm bol}/L_\odot)=-6.60^{+0.03}_{-0.02}$, which is $\approx 15\%$ fainter than the value derived from the retrieved spectrum with inversion.
This suggests that the mechanism responsible for the atmospheric heating in W1935 significantly increases the object's bolometric luminosity by $\approx 15\%$, with the majority of this contribution originating at wavelengths longer than $\approx$5~\um (see Section~\ref{sec:predicted_SED}). 

\subsection{Principal Chemical Species in the Spectra}
We used HITRAN and ExoMol cross sections for various molecular species (CH$_4$, H$_2$O, NH$_3$, H$_2$S, CO, CO$_2$, and PH$_3$) \citep{Gordon_etal2022,Tennyson_etal2016} at a temperature of 475~K (as appropriate for W1935; \citealt{Faherty_etal2024}) and a pressure of 1~bar to identify the principal spectral features in the spectra. 
We found that the dominant features 
shaping the spectra in Figure~\ref{fig:SED} are  CH$_4$, H$_2$O, NH$_3$, CO, and CO$_2$. 
The top panel of Figure~\ref{fig:SED_CF} shows these cross sections throughout the full SED wavelength range. 
As indicated in Figure~\ref{fig:SED}, the $\sim$1--4~$\mu$m near-infrared and 5--20 mid-infrared are dominated by water, methane, and ammonia features, while the 4--5~$\mu$m region exhibit strong carbon-bearing features by CO and CO$_2$. 
The contribution functions from the retrieved models of W1935 are shown in the middle and lower panels of Figure~\ref{fig:SED_CF}. 
The majority of the 1--20~$\mu$m wavelength range probes the pressure range $\approx$60--0.002~bar in the model with inversion and $\approx$60--0.1~bar (with a few wavelength points reaching as low as 0.03 bar) in the model without inversion. The G395H data cover similarly low pressures as the model with inversion but probe shallower layers, down to $\sim$10~bar. The principal molecules identified in the SED  exhibit features in this pressure range, indicating that they may cover different atmospheric layers.

Figure~\ref{fig:SED_aux} facilitates visualization of the pressure levels probed by different wavelengths in the W1935 SED. 
We observe that overall fluxes at the center of prominent absorption bands originate preferentially at higher, lower-pressure layers compared to the fluxes around the corresponding features.
This may be explained by the species' gas opacity that prevents the observation of deeper and hotter atmospheric layers with respect to the pseudo-continuum regions.

\subsection{Predicted near- and mid-infrared SED for W1935}
\label{sec:predicted_SED}
Both retrieved spectra with and without thermal inversion predict similar $\lesssim$2~$\mu$m near-infrared fluxes, however, they exhibit significant differences at specific longer wavelength regions (Figures~\ref{fig:SED} and \ref{fig:SED_CF}). The spectrum with the thermal inversion has weaker $\gtrsim$2~$\mu$m water, methane, and ammonia features, specifically at the $\approx$2.2--2.9~$\micron$, 3.1--4.0~$\mu$m, 5.2--8.4~$\mu$m, and 10.0--11.5~$\mu$m regions. 
Interestingly, the CO and CO$_2$ absorption features at 4.1--5.0~$\mu$m are consistent in both model spectra and with the G395H spectrum. 
This may be explained by the fact that the CO and CO$_2$ feature regions probe atmospheric layers deeper than where the heating occurs, and are therefore not affected, as further discussed below.

The contribution functions in Figure~\ref{fig:SED_CF} and Figure~\ref{fig:SED_aux} show that these CO and CO$_2$ features originate at the highest pressure ($\sim$2--10~bar) layers probed by mid-infrared wavelengths, while the water, methane, and ammonia thermal inversion-sensitive features arise from the lowest pressures ($\sim$0.002--0.04~bar) probed in the mid-infrared. 
We note that the thermal inversion reported in \citet{Faherty_etal2024} occurs at pressures shallower than about 0.3~bar, overlapping with the pressure range probed by mid-infrared water, methane, and ammonia features. 
Thus, the location of the temperature inversion relative to the atmospheric layers probed at different wavelengths explains why H$_2$O, CH$_4$, and NH$_3$ features change, while CO and CO$_2$ features remain unaffected.
Accordingly, the methane emission feature at 3.326~$\mu$m originates in the upper atmospheres, at slightly shallower pressures $\lesssim$0.03~bar).

The model with thermal inversion predicts an additional emission feature at $\approx$7.7~\um (Figure~\ref{fig:SED}), which is absent in the model without inversion. 
The feature appears similar to the methane emission feature at 3.326~$\micron$ observed in the W1935 NIRSpec spectrum and modeled in the retrieved spectrum with atmospheric heating. 
Furthermore, the wavelength regions of both the 3.326~$\micron$ and the 7.7~$\micron$ features probe similar pressures ($\lesssim$0.03~bar) layers (Figure~\ref{fig:SED_CF}; bottom panel).
We therefore predict that W1935 will exhibit more methane emission, particularly in the mid-infrared at $\approx$7.7~\um. 
This prediction is supported by the detection of methane emission at similar wavelengths in Jupiter from both JWST MRS \citep{Rodriguez-Ovalle_etal2024} and Cassini 
Composite Infrared Spectrometer (CIRS)
\citep{Fletcher_etal2009} spectra. 
In addition to methane emission features, our model with inversion also suggests tentative ammonia emission features at $\sim$5.5--6.5~\um and $\sim$10--11~\um, which are weaker than the methane emissions but originate at similarly high altitudes.
The JWST Cycle 4 GO program 7793 (PI: J. Faherty), which will obtain MIRI/MRS and NIRSpec/PRISM spectra of W1935, will allow us to test these mid-infrared predictions for the object and investigate the evolution of the 3.326~$\micron$ emission feature.  
The 40-hour duration of the observations from this upcoming program may also permit constraints on W1935's rotation period via probing for variability in any of its aforementioned spectral features shown in Figure~\ref{fig:SED}.

We note that the emission-like feature at $\sim$6.3~\um in the retrieved spectrum without inversion (brown spectrum in Figure~\ref{fig:SED}) is similar to that produced by the 5.2--7.25~\um water absorption in other cold brown dwarfs \citep[e.g.;][]{Cushing_etal2006,Suarez_Metchev2022}.  
This behavior of the water absorption feature, which appears to split into two features, is caused by a local minimum in water opacity at 6.3~\um.
This allows the observation of deeper, hotter atmospheric layers, resulting in higher fluxes at $\sim$6.3~\um compared to the surrounding wavelengths.

Both retrieved spectra with and without inversion exhibit an additional absorption feature at $\sim$15.5--16.0~$\mu$m. The feature is slightly weaker in the spectrum with atmospheric heating, consistent with the trends seen in the other water, methane, and ammonia mid-infrared absorption features.
This is a relatively unexplored feature and according to to our modeling, it corresponds to NH$_3$. 
\citet{Alejandro_Merchan_etal2025} detected a similar feature at $\sim$14--16~$\mu$m in a Spitzer IRS spectrum of the warmer ($\approx$730~K) T8 dwarf 2MASS J04151954$-$0935066 and tentatively assigned it to a combination of CO$_2$, H$_2$O, and NH$_3$. 
We inspected Spitzer IRS of other objects with similar spectral types, sufficient spectral coverage, and S/N$\gtrsim$10 in \citet{Suarez_Metchev2022}, but none of them exhibit this $\gtrsim$14~$\mu$m feature, highlighting its rarity. 

One of the regions showing the strongest flux differences due to the temperature inversion is the $\sim$5.5--8.5~\um region.  This area is also covered by the IRAC Ch3 and Ch4 filters, which are centered at 5.7~\um and 7.9~\um, respectively. 
In addition, the IRAC Ch1 filter probes the 3.1--3.95~\um methane absorption also sensitive to atmospheric heating. 
Contrarily, the IRAC Ch2 filter (centered at 4.4~\um) that probes the 4.1--5.0~\um CO$_2$ and CO features is essentially unaffected by the thermal inversion, as explained above. 
Thus, mid-infrared color-magnitude diagrams (CMDs) that combine IRAC Ch2 with Ch1, Ch3, or Ch4 might be helpful for identifying objects with upper atmospheric heating and explaining the peculiar position of W1935 in these CMDs (e.g.; Extended Data Figure 1 in \citealt{Faherty_etal2024}). We explore this in the next section.

\subsection{IRAC color-magnitude diagrams}
\label{sec:CMDs}

\begin{figure*}
    \centering
	\includegraphics[width=.32\linewidth]{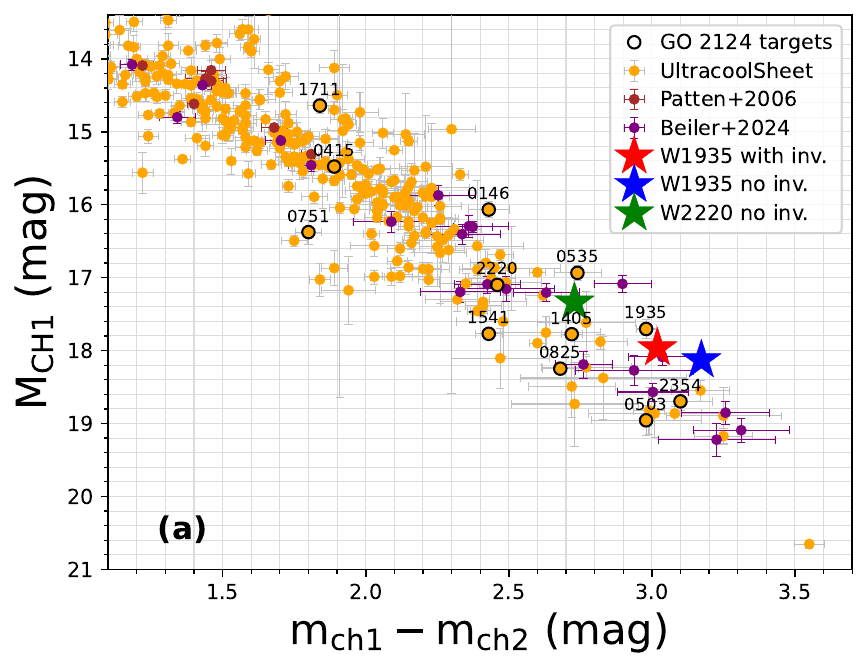}
	\includegraphics[width=.32\linewidth]{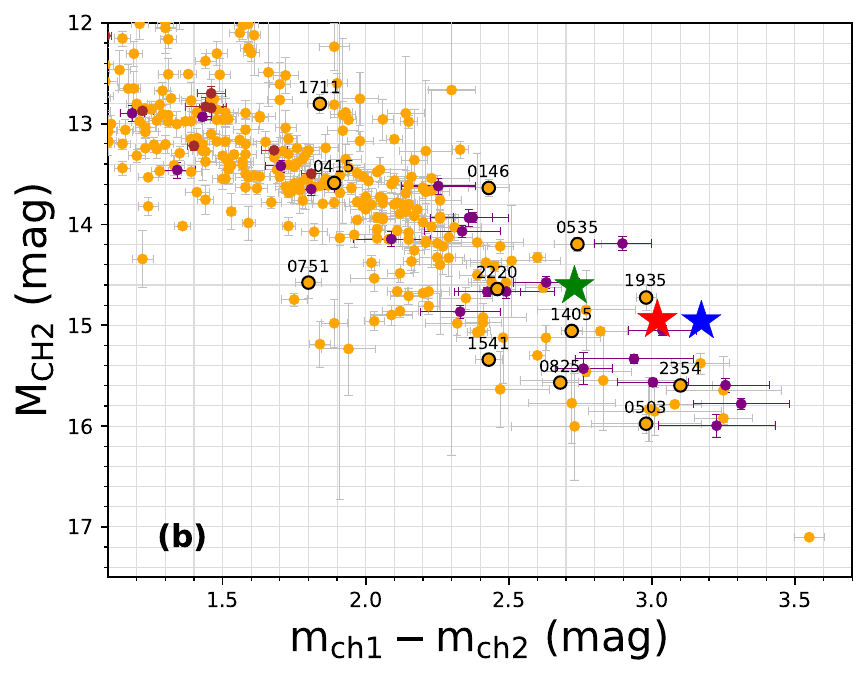}
	\includegraphics[width=.32\linewidth]{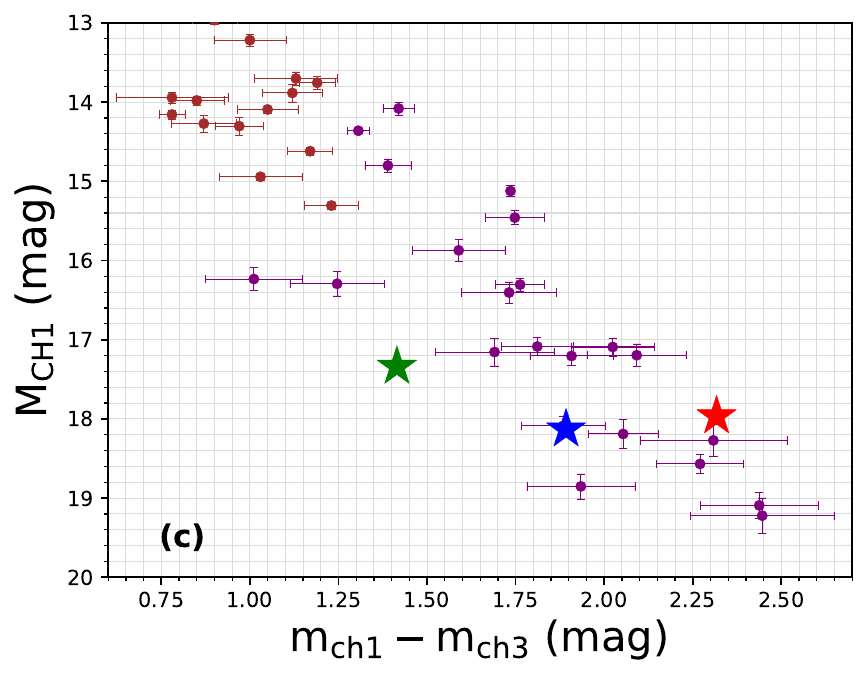}
	\includegraphics[width=.32\linewidth]{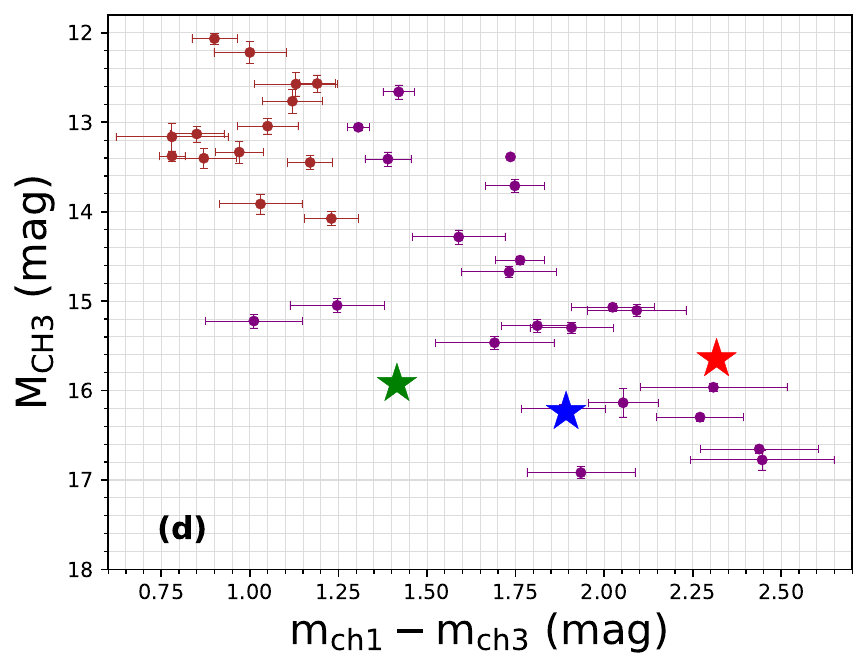}
	\includegraphics[width=.32\linewidth]{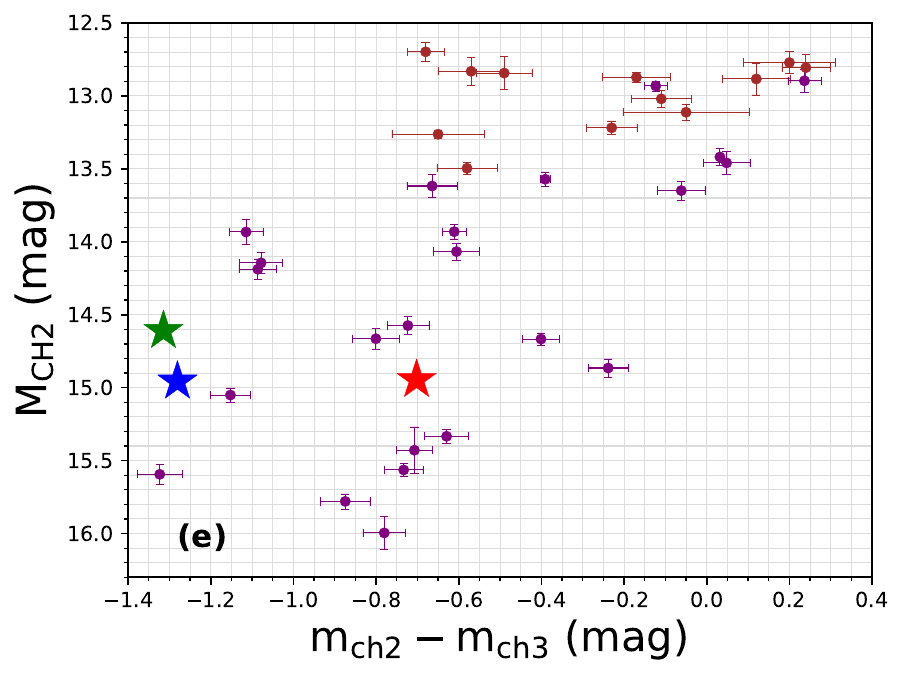}
	\includegraphics[width=.32\linewidth]{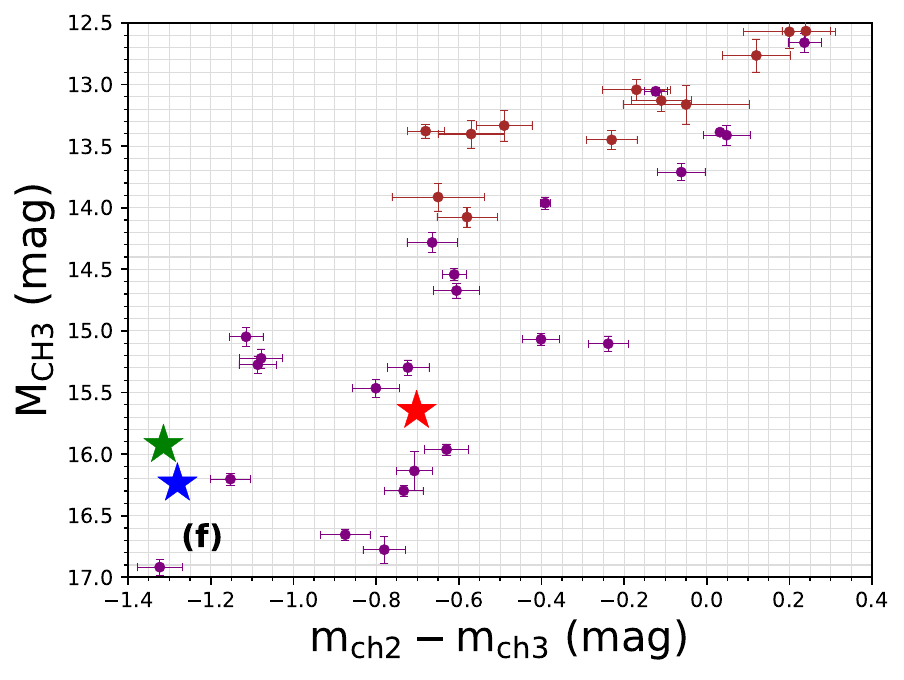}
	\includegraphics[width=.32\linewidth]{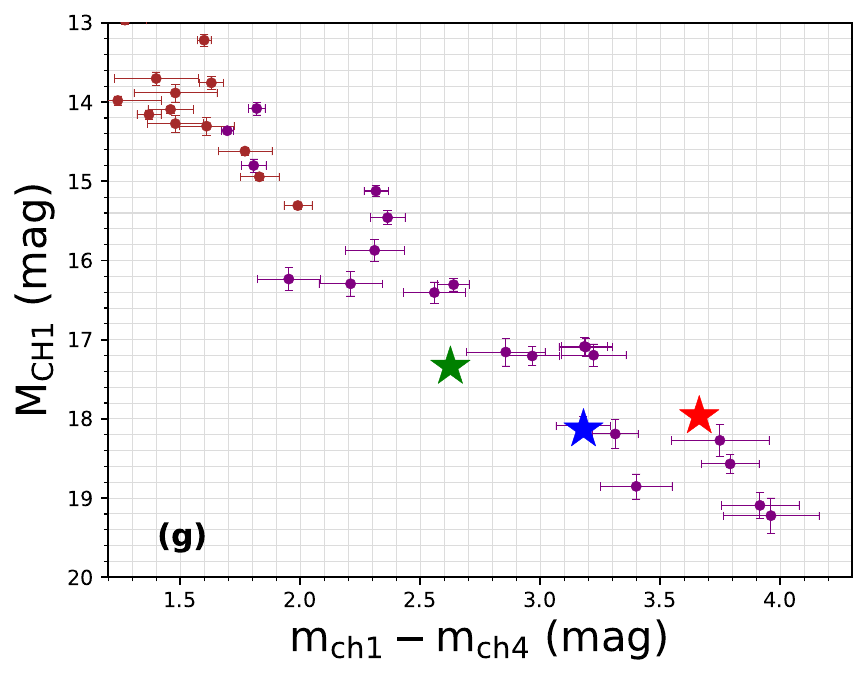}
	\includegraphics[width=.32\linewidth]{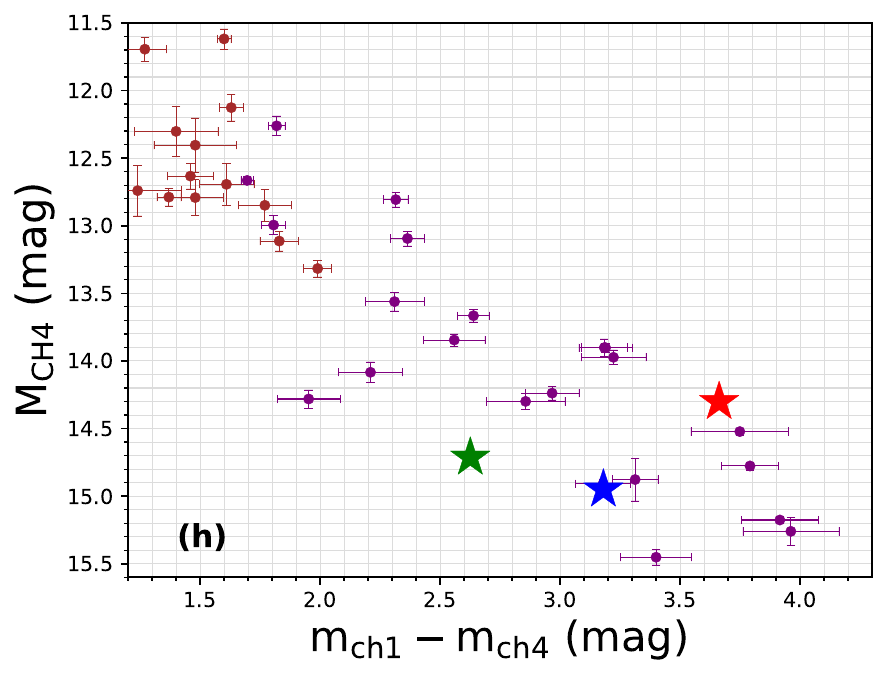}
	\includegraphics[width=.32\linewidth]{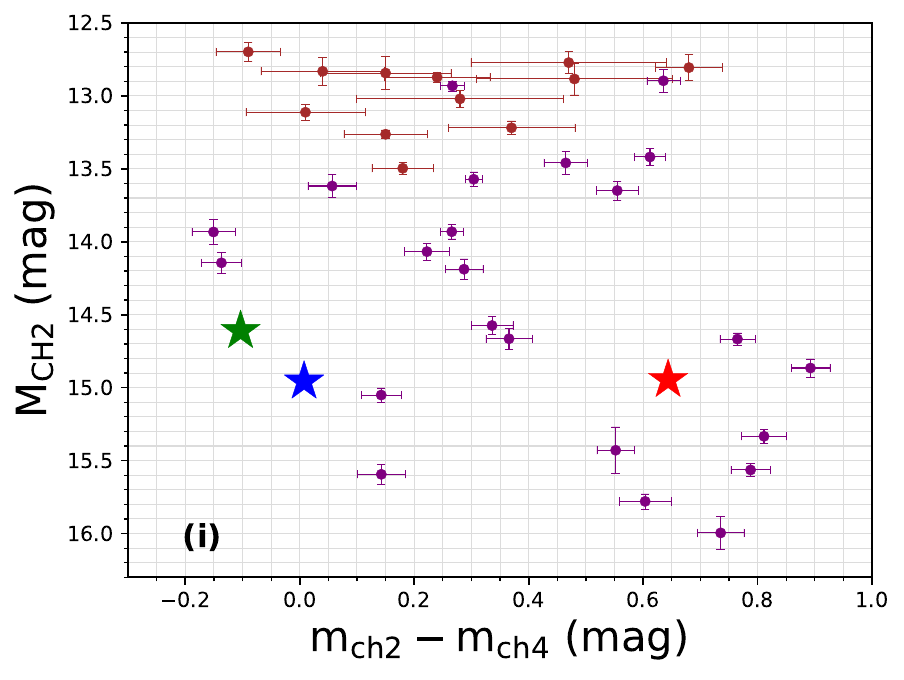}
	\includegraphics[width=.32\linewidth]{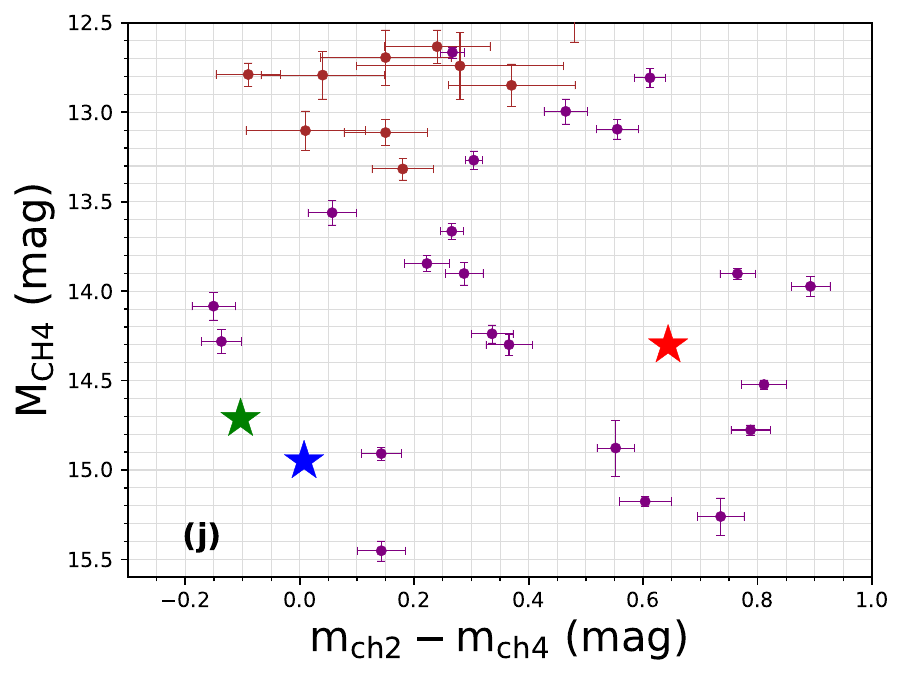}
	\includegraphics[width=.32\linewidth]{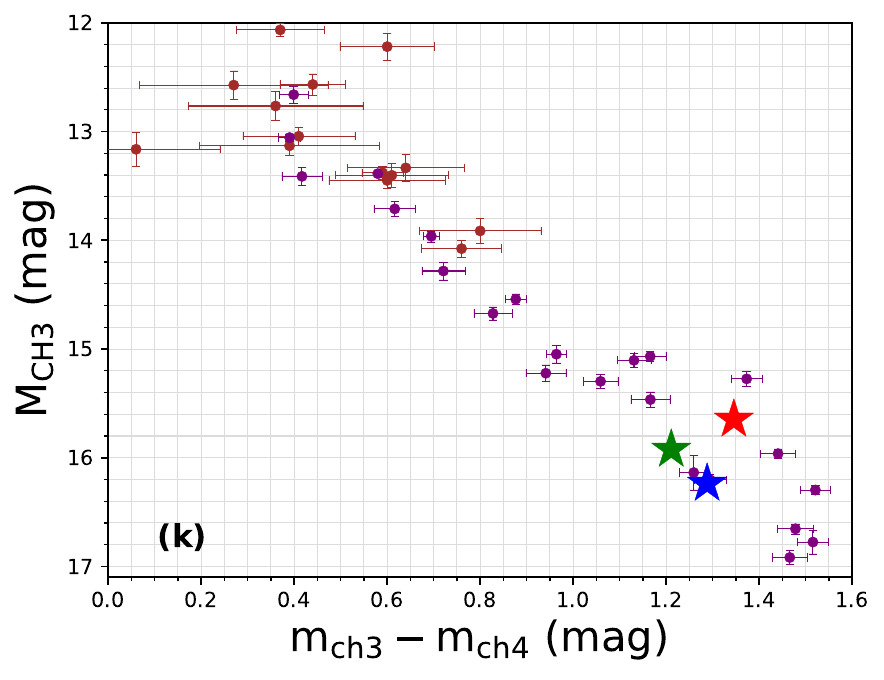}
	\includegraphics[width=.32\linewidth]{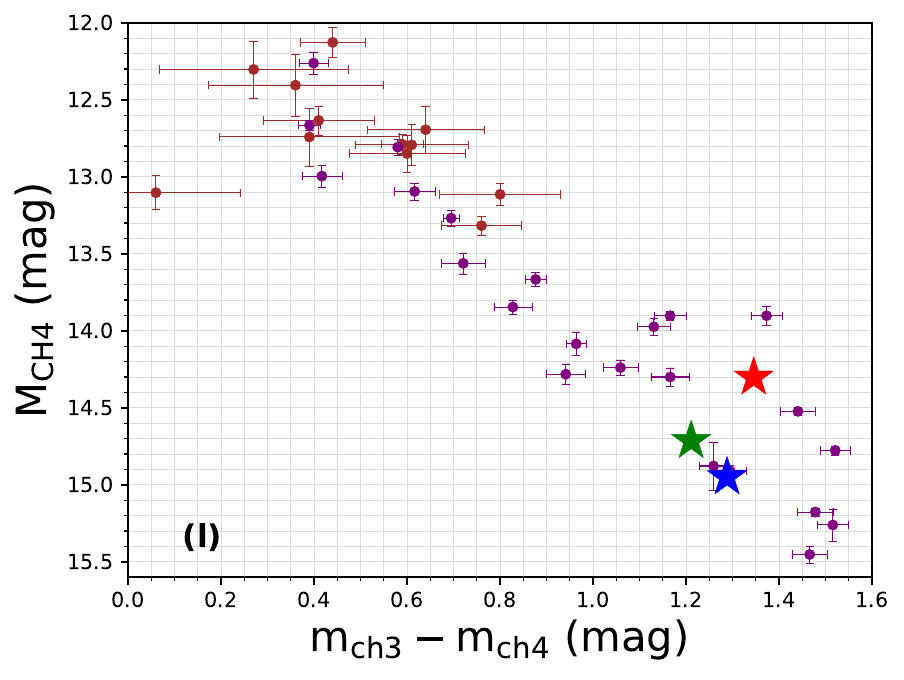}
    \caption{
    CMDs for all combinations of IRAC Ch1, Ch2, Ch3, and Ch4 magnitudes using observed data for $\ge$T5 dwarfs in the UltracoolSheet catalog \citep[][orange points in (a) and (b) panels]{Best_etal2024} and from \citet[][brown points]{Patten_etal2006}, as well as synthetic values for objects with NIRSpec PRISM spectra in \citet[][purple points]{Beiler_etal2024}. The positions of W1935 and W2220 considering synthetic values derived using the retrieved spectra 
    in \citet{Faherty_etal2024} are indicated with a red star (from model with inversion) and a blue star (from model without inversion) for W1935 and a green star for W2220. Labeled objects in (a) and (b) panels indicate the targets in the GO Cycle 1 2124 program.
    }
    \label{fig:CMDs}
\end{figure*}

Figure~\ref{fig:CMDs} shows mid-infrared CMDs for all combinations of IRAC filters: 
$M_{\rm Ch1}$ or $M_{\rm Ch2}$ vs. $m_{\rm Ch1}-m_{\rm Ch2}$ (Figure~\ref{fig:CMDs}a and ~\ref{fig:CMDs}b), 
$M_{\rm Ch1}$ or $M_{\rm Ch3}$ vs. $m_{\rm Ch1}-m_{\rm Ch3}$ (Figure~\ref{fig:CMDs}c and ~\ref{fig:CMDs}d), 
$M_{\rm Ch2}$ or $M_{\rm Ch3}$ vs. $m_{\rm Ch2}-m_{\rm Ch3}$ (Figure~\ref{fig:CMDs}e and ~\ref{fig:CMDs}f),
$M_{\rm Ch1}$ or $M_{\rm Ch4}$ vs. $m_{\rm Ch1}-m_{\rm Ch4}$ (Figure~\ref{fig:CMDs}g and ~\ref{fig:CMDs}h), 
$M_{\rm Ch2}$ or $M_{\rm Ch4}$ vs. $m_{\rm Ch2}-m_{\rm Ch4}$ (Figure~\ref{fig:CMDs}i and ~\ref{fig:CMDs}j), and 
$M_{\rm Ch3}$ or $M_{\rm Ch4}$ vs. $m_{\rm Ch3}-m_{\rm Ch4}$ (Figure~\ref{fig:CMDs}k and ~\ref{fig:CMDs}l). 
All panels include cold brown dwarfs from two samples: \citet[][brown points]{Patten_etal2006} and \citet[][purple points]{Beiler_etal2024}. For the latter sample, we derived synthetic magnitudes from NIRSpec spectra. CMDs that use Ch1 and Ch2 magnitudes also include the larger sample from the UltracoolSheet catalog \citep[][orange points]{Best_etal2024}. 
Additionally, we plot the positions of W1935 and W2220 based on synthetic magnitudes from their retrieved spectra. 
In the following sections, we contextualize the position of W1935 within the larger sample and discuss its peculiarities in these CMDs.

\subsubsection{W1935's Mid-infrared Overluminousity and/or Redness Cannot Be Explained by Atmospheric Heating}
\label{sec:peculiarity}

The synthetic IRAC photometry of W1935 derived from the retrieved spectra with and without inversion shows negligible differences in Ch2 ($\Delta m=-0.003$~mag) compared to the difference of $-0.17$~mag for Ch1 and $-0.6$~mag for Ch3 and Ch4. 
In all cases, the source appears brighter in the spectrum with inversion. 
These differences are produced by the weaker 3.1--4.0~\um and 5.2--8.4~\um water, methane, and ammonia absorption features in the retrieved spectrum with inversion (Figure~\ref{fig:SED}). 
Thus, the overluminous and/or redder than expected position of W1935 compared to other cold brown dwarfs with similar magnitude or color in the $M_{\rm Ch1}$ or $M_{\rm Ch2}$ vs. $m_{\rm Ch1}-m_{\rm Ch2}$ CMDs would be even more pronounced if W1935 lacked a temperature inversion, with the object being 0.17~mag redder in the $m_{\rm Ch1}-m_{\rm Ch2}$ color, as observed in Figure~\ref{fig:CMDs}a and \ref{fig:CMDs}b. 

Given that a thermal inversion in the upper atmosphere leads to brighter Ch1 magnitudes but does not significantly affect Ch2, objects exhibiting this phenomenon are expected to display bluer $m_{\rm Ch1}-m_{\rm Ch2}$ colors compared to those without such inversion. 
Therefore, the overluminous and/or redder than expected position of W1935 in the $M_{\rm Ch1}$ or $M_{\rm Ch2}$ vs. $m_{\rm Ch1}-m_{\rm Ch2}$ CMDs cannot be explained by the atmospheric heating, which leaves open the possibility of other phenomena such as binarity, gravity, metallicity, and/or clouds. 
In fact, MIRI F1000W and F1280W images of W1935 have recently revealed that it is a near-equal mass binary system \citep{DeFurio_etal2025}. 
With this discovery, all four overluminous targets in Ch2 from GO 2124 program (see Methods section of \citealt{Faherty_etal2024}) have been either resolved as binary systems (J014656.66+423410.0; \citealt{Dupuy_etal2015a}, J171104.60+350036.8; \citealt{Liu_M._etal2012}, and W1935; \citealt{DeFurio_etal2025}) or suggested as binaries based on model fits (J053516.80-750024.9; \citealt{Leggett_Tremblin2024}).  Other confirmed T/Y transition binary systems also exhibit brighter Ch2 magnitudes or redder $m_{\rm Ch1}-m_{\rm Ch2}$ colors than expected, suggesting these colors may serve as a potential indicator of multiplicity 
\citep{BardalezGagliuffi_etal2025}. 
On the contrary, objects with $m_{\rm Ch1}-m_{\rm Ch2}$ colors bluer than expected in these CMDs could be candidates for exhibiting thermal inversions, potentially caused by auroral activity. 
However, none of the blue objects in JWST Cycle 1 GO 2124 (Faherty et al. 2025, in prep.) or \citet[][JWST Cycle 1 GO 2302]{Beiler_etal2024} shows methane emission like W1935.
We note, however, that the S/N of the spectra in both programs is very low (median S/N$\approx$1) in the methane emission region (3.30--3.35~\um), increasing up to 15 in \citet{Beiler_etal2024} sample and 4 in the GO 2124 sample.

\subsubsection{W1935 Is an Outlier Only in the Ch1-Ch2 Color}
We observe in Figure~\ref{fig:CMDs} that CMDs constructed with the Ch1, Ch3, and Ch4 filters (Figure~\ref{fig:CMDs}c, \ref{fig:CMDs}d, \ref{fig:CMDs}g, \ref{fig:CMDs}h, \ref{fig:CMDs}k, and \ref{fig:CMDs}l) display well-defined sequences. 
In contrast, the inclusion of Ch2 significantly increases the dispersion (Figure~\ref{fig:CMDs}e, \ref{fig:CMDs}f, \ref{fig:CMDs}i, and \ref{fig:CMDs}j), making it difficult to identify clear sequences, particularly in diagrams involving $m_{\rm Ch2}-m_{\rm Ch4}$. 
The exception is CMDs that incorporate the $m_{\rm Ch1}-m_{\rm Ch2}$ color (Figure~\ref{fig:CMDs}a and \ref{fig:CMDs}b), which still follow clear trends. 
This may be related to the fact that the Ch2 filter probes the disequilibrium chemistry-sensitive features CO$_2$ (4.1--4.4~\um ) and CO (4.5--5.0~\um) \citep{Miles_etal2020}, which do not correlate directly with spectral type or temperature \citep[e.g.;][]{Beiler_etal2024}. Conversely, the CH$_4$, H$_2$O, and NH$_3$ features probed by the Ch1, Ch3, and Ch4 filters strengthen systematically towards later spectral types or cooler temperatures \citep[e.g.;][]{Cushing_etal2006,Suarez_Metchev2022}.

Across the CMDs that trace well-defined sequences, W1935 stands out as an outlier---appearing brighter and/or redder than expected---only when the Ch2 filter is involved. 
In CMDs based on Ch1, Ch3, and Ch4, W1935 aligns more consistently with other objects of similar brightness or color. Thus, the use of Ch2 in CMDs either isolates W1935 as an outlier or increases the overall scatter. 
This could be explained by the fact that the Ch1, Ch3, and Ch4 magnitudes are all affected in the same direction by the thermal inversion-sensitive CH$_4$ and H$_2$O+NH$_3$ features (Figure~\ref{fig:SED}), so their differences hide the peculiarity of W1935. 

\subsubsection{Discrepant Imaging and Synthetic IRAC Ch1 Magnitudes}

\citet{Beiler_etal2024} and \citet{Luhman_etal2024} reported a systematic offset of $\sim$0.3~mag between synthetic IRAC Ch1 magnitudes (derived from NIRSpec spectra) and the corresponding observed values. 
This discrepancy may arise from inaccuracies in the filter response functions \citep[][]{Luhman_etal2024}. 
We found a similar Ch1 offset for W1935 and W2220, with their observed magnitudes \citep{Kirkpatrick_etal2012,Meisner_etal2020} being $\approx$0.25~mag brighter than the synthetic magnitudes derived from the best retrieved spectra. 
However, while W2220 shows a negligible offset ($-$0.03~mag) in Ch2, W1935 exhibits a Ch2 offset (0.21~mag) comparable to its Ch1 offset.
This leads to a larger discrepancy between the predicted and observed positions of W2220 in the $M_{\rm Ch1}$ or $M_{\rm Ch2}$ vs. $m_{\rm Ch1}-m_{\rm Ch2}$ CMDs compared to the corresponding discrepancy for W1935 (Figure~\ref{fig:CMDs}a and \ref{fig:CMDs}b).

\subsection{Forward modeling of the NIRSpec spectrum}
\label{sec:forward_modeling}

\begin{figure*}
	\centering
	\includegraphics[width=0.70\linewidth]{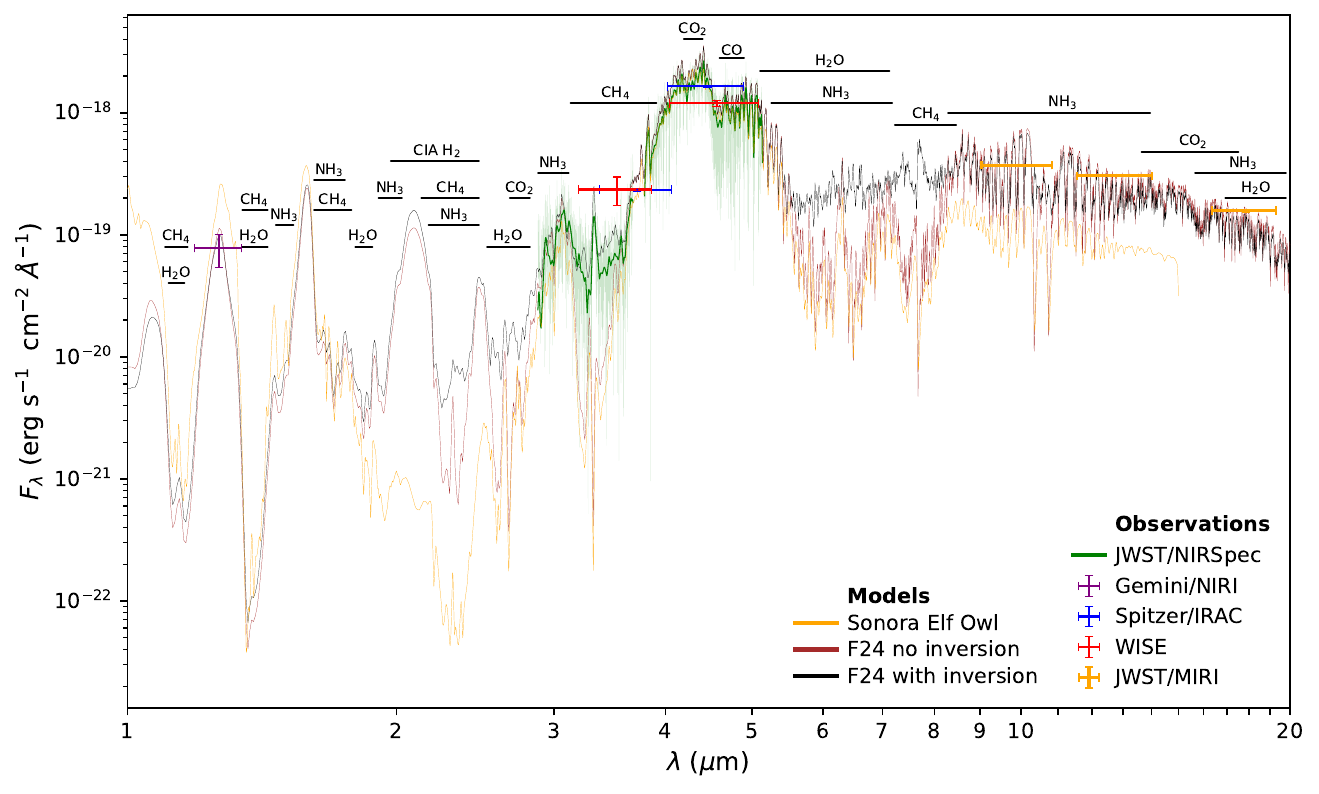}
	\includegraphics[width=0.70\linewidth]{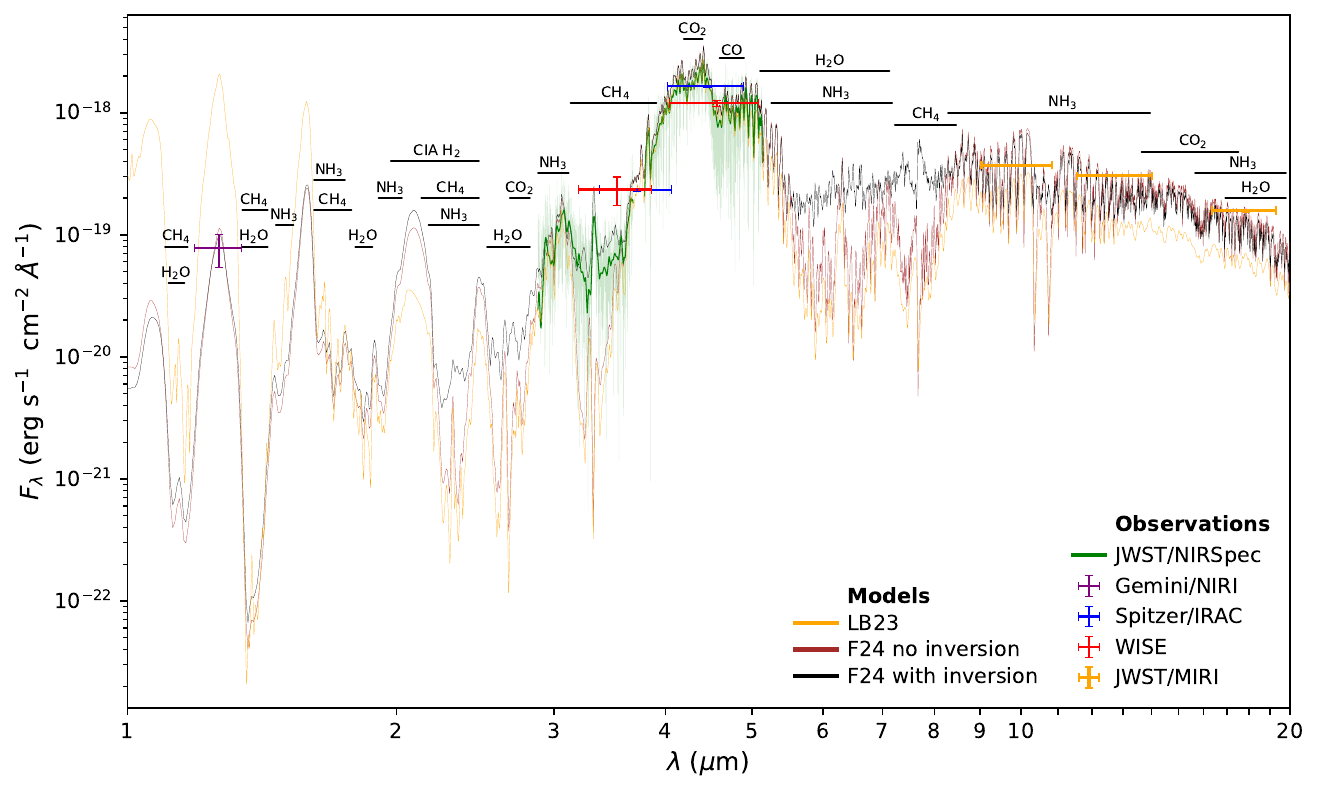}
	\includegraphics[width=0.70\linewidth]{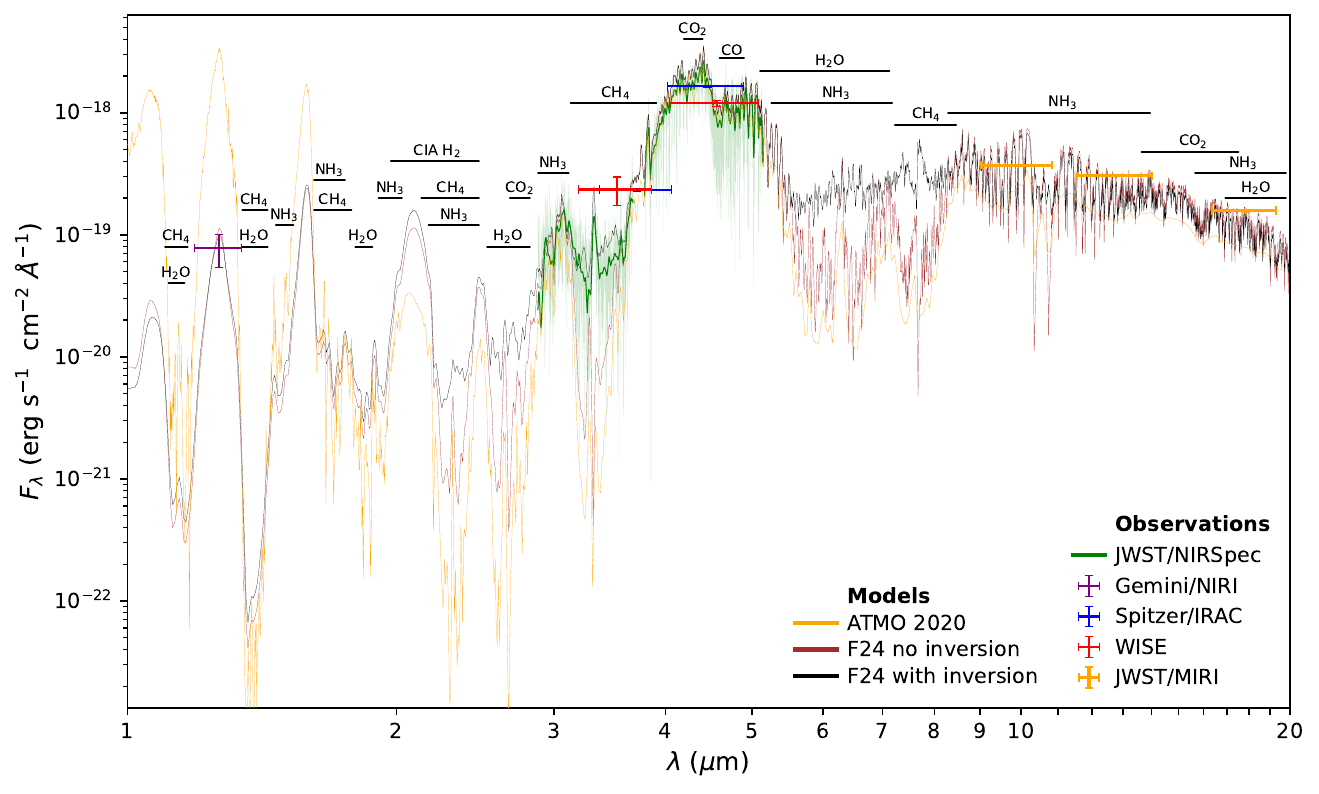}
	\caption{Best self-consistent atmospheric model fits to the W1935 G395H spectrum. Details are the same as Figure~\ref{fig:SED} but including the best fitting-models (dashed orange line) from Sonora Elf Owl (top panel),     
	\citet[][middle panel]{Lacy_Burrows2023}, and ATMO 2020 (bottom panel).} 
	\label{fig:SED_fits}
\end{figure*}

\begin{table}
\begin{flushleft}
\caption{Parameter Coverage of Atmospheric Models.}
  \scriptsize
  \label{tab:models_coverage}
	\begin{tabular}{lcc}
	\toprule
	Parameter & Range & Step or Values \\
	\midrule                                                                      
	\multicolumn{3}{c}{\textbf{Sonora Elf Owl \citep{Mukherjee_etal2024}}} \\
	\midrule                                                                      
	\multirow{3}{*}{\Teff}    & \multirow{3}{*}{275--2400~K} & 25~K (275--600~K)             \\
	                           &                              & 50~K (600--1000~K)            \\
	                           &                              & 100~K (1000--2400~K)          \\
	\logg                     & 3.00--5.50                   & 0.25                          \\
	\logKzz$^a$               & 2--9                         & 2, 4, 7, 8, and 9             \\
	\multirow{2}{*}{$[M/H]^b$}& \multirow{2}{*}{$-1.0$--1.0} & $-1.0$, $-0.5$, 0.0,          \\
	                           &                              & 0.5, 0.7, and 1.0             \\
	(C/O)/(C/O)$_\odot$       & 0.5--2.5                     & 0.5, 1.0, 1.5, 2.5            \\
	\midrule                                                                      
	\multicolumn{3}{c}{\textbf{\citet{Lacy_Burrows2023}}} \\
	\midrule                                                                      
	\multirow{2}{*}{\Teff}    & \multirow{2}{*}{250--800~K}    & 25~K (250--600~K)             \\
	                           &                              & 50~K (600--800~K)             \\
	\logg                     & 3.50--5.00                   & 0.25                          \\
	$[M/H]$                   & $-0.5$--0.5                  & $-0.5$, 0.0, and 0.5          \\
	  \logKzz                   & 6                            & 6                             \\
    \lmix$^c$                 & 0.01--1                      & 0.01, 0.1, and 1              \\
	\midrule                                                                      
	\multicolumn{3}{c}{\textbf{ATMO 2020 \citep{Phillips_etal2020}}} \\
	\midrule                                                                      
	\multirow{2}{*}{\Teff}    & \multirow{2}{*}{200--3000~K} & 50~K (200--600~K)             \\
                              &                              & 100~K (600--3000~K)           \\
	\logg                     & 2.5--5.5                     & 0.5                           \\
	\logKzz                   & 0--6                         & 0, 4, and 6                   \\
	\bottomrule
	\end{tabular}
	\par $^a$Logarithm of the vertical eddy diffusion parameter $K_{zz}$.\\
    \par $^b$Metallicity relative to the Sun.\\
    \par $^c$Mixing length.\\
\end{flushleft}
\end{table} 

To investigate whether current atmospheric models are capable of explaining the observations for W1935 and the prediction by the retrieved spectra, we compared the data to several modern self-consistent atmospheric models using the \texttt{SEDA} (Spectral Energy Distribution Analyzer) python package \citep[][Su\'arez et al. 2025, in prep.]{Suarez_etal2021a}\footnote{\url{https://seda.readthedocs.io/en/latest/}}. 
\texttt{SEDA} uses the \texttt{DYNESTY} dynamic nested sampling \citep{Speagle_2020} for estimating Bayesian posteriors and evidences.  
We consider the Sonora Elf Owl \citep{Mukherjee_etal2024}, \citet[][LB23]{Lacy_Burrows2023}, and ATMO 2020 \citep{Phillips_etal2020} model grids, which are adequate for Y-type objects. 
The free parameters in these models along with their ranges and step sizes are listed in Table~\ref{tab:models_coverage}. 
The three model grids consider disequilibrium chemistry. 
LB23 models also include water clouds. 
Unlike the Sonora Elf Owl and LB23 models, the ATMO 2020 model spectra have a resolution lower than the G395H spectrum resolution ($\sim$250 against $\sim$2700 at 4~$\mu$m), so for ATMO 2020 model comparisons we convolved the G395H spectrum to the resolution of the model spectra.

Figure~\ref{fig:SED_fits} shows the best model fits to the W1935 NIRSpec G395H spectrum from the Sonora Elf Owl (top panel), LB23 (middle panel), and ATMO 2020 (bottom panel) models. 
The parameter constrains from the forward modeling of the G395H spectrum are listed in Table~.
The best-fitting model from each grid was convolved to a resolution of $R\sim250$ at 4~\um.
In general, models reproduce similarly well the 4.3~$\mu$m CO$_2$ and 4.75~$\mu$m CO absorption features in the G395H spectrum, but fail to explain the $\sim$3.1--3.9 methane feature in the G395H spectrum, which includes the 3.326~$\mu$m methane emission feature. 
The predicted $\sim$3.1--3.9\um methane feature by the models is significantly stronger than the observation, but the pseudo-continuum regions around the feature agree. 

We also judge how the best self-consistent model fits to the W1935 G395H spectrum compared to the retrieved spectra across the entire 1--20~\um wavelength range. 
None of the self-consistent models reproduces 
either the $\sim$3.1--3.9~\um and 7.1--8.6~\um methane or the 5.2--7.0~\um H$_2$O+NH$_3$ absorption features in the retrieved spectrum with a temperature inversion. 
The self-consistent model spectra exhibit stronger features so, as expected, they are a better fit to the retrieved spectrum without inversion in the mentioned wavelength ranges. 
However, unlike the other models, the best-fitting Sonora Elf Owl model predicts significantly fainter fluxes beyond $\approx$8~$\mu$m and at the $K$-band compared to the retrieved spectra or the other self-consistent model fits. 
The underlying cause of these differences is unclear, and we anticipate similar issues in the updated Sonora Elf Owl model grid \citep{Wogan_etal2025}, since it only corrects the CO$_2$ concentrations, which are not the dominant species at such long wavelengths. 
All self-consistent model spectra also show intriguing differences in the near-infrared. Both retrieved spectra predict a $J_{\rm MKO}$ magnitude that is consistent within 2--3$\sigma$ of the value reported by \citet{Leggett_etal2021}, as shown in Figure~\ref{fig:SED}. 
However, the best-fitting self-consistent models exhibit significantly stronger and brighter near-infrared spectral features. 

Although we discuss the performance of the self-consistent model fits to the relatively narrow G395H spectrum range (2.9--5.1~\um) across a much broader wavelength span (1--20~\um), it is likely that these models better predict the near-infrared observations of cold brown dwarfs when compared in that specific range or across a broader wavelength range than the one covered by NIRSpec G395H. 
Nevertheless, the mid-infrared discrepancies between the self-consistent models and G395H spectrum reinforce the need for detailed modeling, such as retrieval analysis \citep[e.g.; ][]{Hood_etal2024,Faherty_etal2024}, to successfully capture all the details in the observations of cold brown dwarfs.

\section{Summary and Conclusions}
\label{sec:conclusions}

Recent analysis by \citet{Faherty_etal2024} revealed that the $\ge$Y1, $T_{\rm eff}\approx$482~K dwarf W1935 exhibits methane emission in its NIRSpec G395H spectrum. 
Modeling by the authors using the retrieval technique indicates heating in the upper atmosphere, likely driven by auroral activity. 
We present here an extension of the retrieved spectra to the G395H spectrum beyond the G395H wavelength coverage to both shorter and longer wavelengths, covering the full 1--20~\um range. 
By comparing the retrieved spectra with and without thermal inversion---both against each other and with self-consistent atmospheric models---and by inspecting IRAC CMDs, we found that:

\begin{itemize}
    \item The atmospheric heating in W1935 increases its bolometric luminosity by $\approx$15\% according to the model prediction with vs. without a temperature inversion (Section~\ref{sec:Lbol_change}). 
    \item The upper atmospheric heating affects the spectral features originating at lower pressures (higher altitudes) than the zone where the heating occurs ($\lesssim$0.3~bar)(Figure~\ref{fig:SED_CF}). 
    In particular, wavelengths shorter than $\approx$2~\um, as well as the 4.1--5.0~\um CO$_2$ and CO spectral signatures, are not significantly altered by the temperature inversion. However, water, methane, and ammonia absorption features at longer wavelengths are substantially weakened (Figure~\ref{fig:SED}). 
    \item The model with thermal inversion predicts an additional methane emission feature at $\approx$7.7~\um, originating from similar pressure layers ($\lesssim0.03$~bar) as the observed emission feature at 3.326~\um in the G395H spectrum (Figures~\ref{fig:SED} and \ref{fig:SED_aux}). This model also suggests tentative ammonia emission features at $\sim$5.5--6.5~\um and $\sim$10--11~\um. 
    \item The peculiar overluminous and/or redder position of W1935 in $M_{\rm Ch1}$ or $M_{\rm Ch2}$ vs. $m_{\rm Ch1}-m_{\rm Ch2}$ CMDs (Figure~\ref{fig:CMDs}a and \ref{fig:CMDs}b) cannot be explained by the atmospheric heating. In fact, the heating tends to produce bluer than expected $m_{\rm Ch1}-m_{\rm Ch2}$ colors. This leaves open the possibility of other phenomena such as binarity to explain this W1935 peculiarity \citep{DeFurio_etal2025}.
    \item There is a $-0.25$~mag residual between the imaging and synthetic IRAC Ch1 magnitudes of W1935 and W2220, which is consistent with the systematic Ch1 offset reported in \citet{Beiler_etal2024} and \citet{Luhman_etal2024} for a number of cold brown dwarfs. 
    \item W1935 occupies an average position in CMDs that combine IRAC filters sensitive to auroral activity, namely Ch1, Ch3, and Ch4 (Figure~\ref{fig:CMDs}). However, when Ch2 is included, either W1935 emerges as a peculiar object (in $m_{\rm Ch1}-m_{\rm Ch2}$ colors) or the dispersion in the CMDs increases significantly (in $m_{\rm Ch2}-m_{\rm Ch3}$ and $m_{\rm Ch2}-m_{\rm Ch4}$ colors), likely due to the fact that the 4.1--5.0~\um CO$_2$ and CO absorption features in cold (late-T and Y type) brown dwarfs do not correlate well with spectral type \citep[e.g.;][]{Beiler_etal2024}.
    \item Atmospheric models can reasonably predict mid-infrared wavelengths that are not affected by the thermal inversion but, as expected, they fail to reproduce the thermal inversion-sensitive CH$_4$, H$_2$, and NH$_3$ features (Figure~\ref{fig:SED_fits}). However, in the absence of near-infrared observations, it is hard to assess in detail the performance of atmospheric models in the case of W1935.
\end{itemize}

The JWST spectrum presented in this article were obtained from the Mikulski Archive for Space Telescopes (MAST) at the Space Telescope Science Institute. The specific observations analyzed can be accessed via \dataset[doi: 10.17909/0vbh-qv13]{https://doi.org/10.17909/0vbh-qv13}.

\section*{Acknowledgements}


\bibliography{mybib_Suarez}
\bibliographystyle{aasjournal}



\end{document}